%% file: paper.tex
\documentstyle[epsfig,preprint,aps,prb]{revtex}
\tightenlines

\renewcommand{\appendix}{%
 \setcounter{section}{0}%
 \setcounter{equation}{0}%
 \renewcommand{\thesection}{{APPENDIX} \Alph{section}}
 \renewcommand{\theequation}{\Alph{section}.\arabic{equation}}%
}

\begin{document}
\preprint{LA-UR 00-306}
\input zabst.tex
\input sect1.tex

\input zref.tex
\input zfigs.tex

\input zfigps.tex
\end{document}

%% file: zabst.tex
\title{\bf Physics of polymer melts: A novel perspective}
\author{Shirish M. Chitanvis}
\address{
Theoretical Division, 
Los Alamos National Laboratory\\
Los Alamos, New Mexico\ \ 87545\\}

\date{\today}
\maketitle
\begin{abstract}
We have mapped the physics of
polymer melts onto a time-dependent Landau-Ginzburg
$\vert\psi\vert^4$ field theory using techniques of functional
integration. 
Time in the theory is simply a label for the location of a given
monomer along the extent of a flexible chain.
With this model, one can show that the limit of infinitesimal
concentration of a polymer melt corresponds to a {\em dynamic} critical
phenomenon.
The transition to the entangled state is also shown to be a
critical point. 
For larger concentrations, when the role of fluctuations is reduced, a 
mean field approximation is justifiably employed to show the existence 
of tube-like structures reminiscent of Edwards' model.  
\end{abstract}


%% file: sect1.tex
Issues in polymer melt physics have continued to provide an enduring
source of theoretical investigations, owing to the complexity of the field.
The basic physics of flexible polymer melts is embodied in Flory's
theorem\cite{deG}. 
This theorem states that for low densities, fluctuations in the
melt are quite pronounced, while for large densities, fluctuations 
are so suppressed that the polymers behave as independent Gaussian
chains again.
One purpose of this paper is to modify the current perspective
of polymer melt physics in the interesting low density regime.
Rather than this state being close
to a critical point in the static sense\cite{deG}, the connectivity of the
chains in the system implies that the physics can be mapped
onto a dynamic critical phenomenon.
In the old picture, global properties such as the radius of
gyration $R_g$ (end-to-end distance) were shown to possess scaling
properties, while local properties of chains could not be addressed. 
The new view-point allows a computation of the
local structure of individual chains as well.
Our estimate of the Flory exponent $\nu$ is approximately
$0.631$, in line with universality arguments, and differs from
the usual estimate of $0.588$.  
More generally, the paper provides a powerful new method to
study polymer physics.
As an example we show that it can be used to describe the approach to
entanglement as yet another critical point.
This transition is beyond the scope of the standard tube model, a mean
field approximation. 

The self-consistent field theory (SCF) discussed by de Gennes\cite{deG} 
is a mean field approximation to study melts of flexible polymers.
Generalizing this model beyond its mean field roots has obvious advantages. 
Such a generalization has been attempted\cite{larad}.
However, while this generalization was appropriate for the
intended application, it
ignored an important property of polymers.
The theory involves a description of polymers in terms of a field
$\psi(\vec r)$, where $\vec r$ is the location 
in physical space of any segment, such that $\vert \psi(\vec
r)\vert^2$ is the probability of finding a segment at $\vec r$.
If the polymer is $N$ segments long, there is no representation in 
this theory of which segment ($1$ through $N$) this field
refers to.
In other words, reference to the connectivity of the chains is missing.
This can be achieved using ideas from functional integration.

The propagator for a single flexible chain may be represented
by\cite{doi-ed}: 

\begin{eqnarray}
G_0(1,2;n) &&\equiv \langle 1,n\vert \left[ \partial_{n} - \left({b^2
      \over 6}\right) 
  \nabla^2 \right]^{-1} \vert 2,0 \rangle \sim  \nonumber\\
&& \int_{\vec R_2} ^{\vec R_1} {\cal D}\vec R(n')
\exp-\left[\left({3\over 2 b^2}\right)  \int_0 ^{n} dn' \left({\partial
      \vec R(n') \over \partial n' }^2 \right)\right]
\label{one}
\end{eqnarray}

where $b$ is the bond length of the polymer, and where $\partial_n
\equiv {\partial \over \partial n}$.
This expression is obtained by considering only the entropy of a
flexible chain.

Alternatively, one knows from methods in functional integration
that\cite{kaku}:

\begin{eqnarray}
&& \langle 1,n\vert \left[ \partial_{n} - \left({b^2\over6}\right)  
  \nabla^2 \right]^{-1} \vert 2,0 \rangle ~\sim~ \int {\cal D}^2\psi
~\psi^*(\vec R_1,n) 
\psi(\vec R_2,0)  \exp-[\beta {\cal F}] \nonumber\\  
&&\beta {\cal F} = \int dn' 
d^3x~\psi^*(\vec x,n') \left[{\partial_{n'}} 
  -\left({b^2\over6}\right) \nabla^2 \right] \psi(\vec x,n')
\label{map}
\end{eqnarray}
where ${\cal D}^2 \psi \equiv {\cal D} \psi^* {\cal D} \psi$,
$ \beta = {1 \over k_B T} $,
$k_B$ is Boltzmann's constant and $T$ is the temperature.
Thus we have another way of thinking about a system of
flexible polymers, in terms of $\psi(\vec x,n)$ and an energy
functional $\beta {\cal F}$ which is isomorphic to one that describes
diffusion.  
Here $(\vec x,n)$ labels the location $\vec x$ in physical
space, of the $n$-th segment of a chain,
and $\vert \psi(\vec x,n)\vert^2$ is the probability of finding a polymer
segment at a given location in space, as suggested by
Eqn.(\ref{conti}) below.
In this sense we have a density functional theory in the style of Kohn 
and Sham.
Kleinert\cite{hklein} and Semenov et al\cite{semenov} have proposed
similar formalisms. 
However this paper uses the new formalism to
probe the physics of polymer melts in a broader sense.

It is possible to derive from the partition function 
${\cal Z}=\int {\cal D}^2 \psi~ \exp-[\beta {\cal F}] $, a $2p$-point
correlation function, which decouples at the non-interacting level
into a product of Green's functions for $p$ independent polymers.
The main advantage of the functional path integral formalism is that
one can  
now model more easily interactions in systems with large numbers of
polymers.
The following model, written to look like a Kohn-Sham type density
functional theory describes excluded volume effects:

\begin{eqnarray}
&&H_0 \to H = H_0 + V - \mu \nonumber\\
&&H_0 \equiv -\left({b^2\over6}\right) \nabla^2 \nonumber\\
&&V = {v\over 2} \vert \psi(\vec x,n)\vert^2
\label{int}
\end{eqnarray}

where $v$ is the usual excluded volume interaction
parameter\cite{doi-ed}, and $\mu$ is the chemical potential invoked in 
the form of a Lagrange multiplier to conserve the number of polymer
segments in the system.

The self-avoiding walk of a solitary
chain can be modeled by starting from
Eqn.(\ref{one})\cite{decloiz}, by adding to the argument of the
exponential on the 
right hand side, a series of terms which describe the excluded volume
interaction between polymer segments.
Caution must be used to apply this approach to a
system of many polymers.

To understand the physics in the $|\psi|^4$ model, let us
extremize the functional density ${\cal F}$, while considering only
solutions homogeneous in space and in $n$.
This yields a maximum at $\psi = 0$, and minima at:

\begin{equation}
\vert \psi \vert^2 = {\mu\over v}
\label{min}
\end{equation}

This of course leads to an infinity of solutions, equivalent within a
phase difference.
Following quantum field theory ideas, the phase of the field can be
related to scattering, or interaction effects.
The physical system can be thought of as localized regions
where Eqn.(\ref{min}) is satisfied, separated by domain walls which
permit the transition from one minimum to another.

Note that Eqn.(\ref{min}) is equivalent to an estimate of the chemical 
potential ($\mu_0 = c_0 v$) if the average number density $c_0$ is known.
From this view-point, $c_0 \to 0$ represents a system of polymers 
which approaches a critical point from {\em below}.
Here we have in mind an analogy with the usual Landau-Ginzburg
$\phi^4$ model, where the vanishing of the coefficient of the
quadratic term in the energy functional leads to a single well
potential, signifying a critical point in the phase diagram.
This issue was treated by de Gennes
by mapping the polymer problem onto one in a zero-component $\phi_j^4$ 
field theory (the self-avoiding walk of a solitary
chain)\cite{decloiz}. 
The different perspective offered by our theory is that
the physics of low concentration melts is really an issue in {\em
  dynamic} critical phenomena, given the  
degree of freedom represented by the variable $n$.

We can compute a dynamic correction to the free
particle Green's function, from the so-called Saturn 
diagram\cite{binney,hoh,ramond}.
This diagram is the lowest order
non-vanishing $\vec k,~\omega$ dependent contribution to
the self-energy.
To do this calculation, it is first convenient to write $\beta {\cal
  F}$  in a dimensionless form, using $\psi \to \psi/\sqrt 
c_0$, 
and scaling all length scales by $c_0^{-1/3}$.
This yields $v \to \alpha = c_0 v$.
In the limit of small $\mu_0$,

\begin{eqnarray}
\hat G_0(\vec k,\omega) &&\to \left( G_0^{-1}(\vec 
  k,\omega)-\Sigma(\vec k,\omega)  
\right)^{-1} \nonumber\\  
\Sigma(\vec k,\omega) &&\approx 48 \alpha^2 \int {d^3k_1\over (2
  \pi)^3} \int  
{d^3k_2\over (2 \pi)^3} {1 \over {-i \omega + (c_0^{-2/3} b^2/6) (\vec 
    k_1^2 + \vec k_2^2)  
    + (c_0^{-2/3} b^2/6) (\vec k - \vec k_1 - \vec k_2)^2 - 3 \mu_0}} 
\label{sat1}
\end{eqnarray}

where $\hat G_0(\vec k,\omega)$ is the Fourier transform of the 
Green's function introduced in Eqn.(\ref{one}), and
where we have performed the frequency integrations involved in the
diagram using the method of residues, and used the mean field value
for the chemical potential $\mu_0 = c_0 v \to 0$.
This integral can be found in Hohenberg and
Halperin\cite{hoh} and they use Wilson's renormalization scheme to
analyze the properties of this integral.
We have been able to 
evaluate this multi-dimensional integral analytically.  
The integrals were performed using the identity $1/t = \int_0 ^{\infty} 
d\lambda \exp(- \lambda t)$, and introducing a change of variables to
the center-of-mass and relative co-ordinates of $\vec k_1, \vec k_2$.
This allows a separation of variables to occur, permitting an
integration over the momentum variables.
The final
integration is performed using the identity $\int_0 ^{\infty} \exp(-A 
x) x^k dx = A^{1+k} \Gamma(1+k)$.
In using this identity, we have to perform the integration
first in arbitrary dimension $d$, where the identity is valid, and
then 
continue it formally to $d=3-\epsilon,~ \epsilon\to 0$.
The infinities in the integral are then isolated into the Gamma
function.
As usual, the infinite part is removed by introducing an
appropriate counter term in $\beta {\cal F}$.
This amounts to a renormalization of $b$.
The finite result is:

\begin{eqnarray}
\Sigma(\vec k,\omega) &&= \alpha^2 A (2 m)^{3-1/C} \left(
    -i\omega + k^2/(6 m) - 3 \mu_0 \right)^2 \left(C - \ln [ 2 m 
    (-i\omega + k^2/(6 m) - 3 \mu_0 ) ]\right) \nonumber\\ 
C &&= \ln [ 4 \sqrt 3 \pi] + {\rm diGamma}(3) \approx 4 \nonumber\\  
\Sigma(\vec k,\omega) &&\approx \alpha^2 A C (2 m)^{3-1/C} \left( 
    -i\omega + k^2/(6 m) - 3 \mu_0 \right)^{2-1/C} 
\label{self1}
\end{eqnarray}
where 
$A = 2 \sqrt 3 \pi^{-3/2}$, and 
$m = {3 \over c_0^{2/3} b^2}$.
The last approximation in Eqn.(\ref{self1}) applies when the argument
of the 
logarithm has a magnitude less than one.
Since $\mu_0 \le 1$, the approximate scaling form thus holds for
$|\omega| \le 1$ and $k \le \sqrt{18}/b $, with an error of a few
percent or less.
In writing down these equations, we have implicitly 
performed {\em mass} renormalization\cite{hoh}.
The effect of the self-energy $\Sigma$ is more 
pronounced at small length scales than it is at long wavelengths.
The renormalized Green's function can also be written for
$|\vec k| \to 0 $ as:

\begin{equation}
\hat G (\vec k,\omega) \approx {1\over {-i \omega+
    \left({B^2(\omega,\mu_0)\over 6}\right) k^2  - \tilde
    \mu(\omega,\mu_0)} }
\label{phys1}
\end{equation}
where

\begin{eqnarray}
\tilde \mu(\omega, \mu_0) &&= \mu_0 - \alpha^2 A (2 m)^{3-1/C} 
(-i \omega - 3 \mu_0)^{2-1/c} 
\nonumber\\  
B^2(\omega,\mu_0) &&= b^2 c_0^{2/3}+ \Delta b^2(\omega,\mu_0) 
\nonumber\\  
\Delta b^2(\omega,\mu_0) &&= - \alpha^2 (6 A) (2 m)^{3-1/C}~(2-1/C) 
\left( -i \omega - 3 \mu_0 \right)^{1-1/C} 
\label{phys2}
\end{eqnarray}
Note that this approximation holds in the regime discussed below
Eqn.(\ref{self1}). 
Now $-i \omega \equiv{\partial/\partial n}\sim 1/\vert p-q \vert$,
with $p,q$ referring to the 
$p^{th}$ and $q^{th}$ segments respectively, while $k$ is the inverse
separation in physical space of these segments.
In the vicinity of the critical point $\mu_0 = 0$, $\Delta b^2$
displays a scaling
property viz., $\sim -\vert p-q\vert^{-\sigma}$  in the 
appropriate regime,  
where we have defined a scaling exponent $\sigma$  

\begin{equation}
\sigma = (1-1/C) \sim 0.75
\label{p}
\end{equation}
In this regime, the coefficient of the scaling term
behaves as $\sim \alpha^{0.2}$, a weak dependence, if we estimate
$v\sim b^3$. 

It follows from Eqn.(\ref{phys2}) that the effect of the
gradient-smoothing term $\nabla^2$ 
in the energy functional is reduced by an amount 
$-|\Delta b^2|$, and this effect is dominant for segments separated by 
a relatively short distance, as it vanishes in the limit of infinite
separation.
It means that if segments on a chain happen to be in
close proximity in the melt, they will tend to stay together
due to the reduction of the smoothing term.
Neutron scattering experiments or numerical simulations may provide
verification of this notion. 
Neutron experiments are customarily employed in the long wavelength
regime to 
investigate quantities like the radius of gyration of polymeric
systems.
Other techniques, such as those used by Smith et al\cite{smith} may be 
more useful.

The chain takes on the appearance of pearls on a string which push
the ends of the string further away from each other.
One way to calculate the scaling properties of the radius of gyration
within the current model is to
compute vertex corrections within the current model, and then
implement Wilson's scaling arguments.
However, it is easier to go back to the original theory defined by 
the energy density $\psi^* (\partial_n + H) \psi$, and truncate it with
$\partial_n \to 1/N$ where $N$ is the average chain length of the
polymeric system.
This is justified on the grounds that we are interested only in long
range fluctuations.
Accordingly, we also have to restrict the $n'$ integration in $\beta
{\cal F}$ to a small neighborhood around $N$.
After performing this renormalization,
the entire machinery of the static theory of critical phenomena
applies.
And it follows that very near the critical point $\mu_0
= 0^+$, the correlation length $\xi \equiv R_g \sim N^{\nu}$, where
the lowest order field field theoretic techniques yield $\nu \approx
0.6$, the value obtained by Flory.

More accurate calculations yield\cite{doi-ed} a value for $\nu$ closer to
$0.631$\cite{hklein,binney}. 
The standard model for examining the effects of excluded volume
interactions yields\cite{zinn} $\nu \approx 0.588$.
The difference between this model and ours was discussed below
Eqn.(\ref{int}).

There is a transition to an entangled 
state at a value of the concentration $c_0 < 1/v$.
To see this let us begin by computing vertex 
corrections, which can be thought of as giving rise to an effective
coupling constant. 
The lowest order correction comes  
from the so-called fish diagram with two internal lines.
The frequency integral involved in the calculation is performed by
closing the contour in the lower half-plane.
The remaining momentum integral contains a divergent part which
involves an integral from $\sqrt {2 m \mu_0}$ to $\infty$.
This is written as an integral from $0$ to $\infty$ minus a finite part
from $0$ to $\sqrt {2 m \mu_0}$.
The infinite part is removed in the usual manner with a counter term,
and amounts to a renormalization of the coupling constant.
This leads to a new coupling constant $\alpha_R$:

\begin{eqnarray}
\alpha_R(q) &&= \alpha - \alpha^2 \tilde \Gamma (q)
\nonumber\\    
\tilde \Gamma (q) &&\approx \left({(6 \mu_0)^{3/2} \over 
  24 \pi^2 c_0 b^3}\right)~{1\over -i\omega_q -2 \mu_0 +\vec
q^2/2m} +{\cal O}(\vec q^2) 
\label{vertex} 
\end{eqnarray}  
where $q\equiv(\vec q,\omega_q)$, etc. and $\mu_0 <<1$.
Notice that to this order in perturbation theory, an increasing
concentration signified by an increasing $\alpha$ leads to a lower
effective coupling constant, consistent with the latter half of
Flory's theorem.
We can define a beta function as is done conventionally in
Renormalization Group theory, viz.,

\begin{eqnarray}  
\beta(\hat \alpha_R) &&= {\partial \hat \alpha_R 
  \over \partial \ln y} \approx -\hat \alpha_R +\hat \alpha_R^2 \nonumber\\
y &&= -i \omega_q - 2 \mu_0 \equiv 1/N - 2 \mu_0 \nonumber\\ 
\hat \alpha_R &&\equiv \left({(6 \mu_0)^{3/2} \over 
  24 \pi^2 c_0 b^3}\right) \alpha_R(\vec q=0)
\label{FP1}
\end{eqnarray}
where $y$ plays the role of a scaling parameter, and as $y\to 0$
($N\to (2\mu_0)^{-1}$), we
see
that $\hat \alpha_R$ flows towards a nontrivial fixed point
viz., $\hat \alpha_R=1$.
As an example, it can be checked in the case of polydimethylsiloxane
(PDMS), that 
$c_0\sim 3 \times 10^{-3}~cm^{-3}$, $b\sim 1.5\AA$, yielding
$N_c\sim 167$, which is close to the experimentally determined\cite{pahl}
critical entanglement chain length of $\sim 200$.
Similarly, for polystyrene, using $b\sim 1\AA$, we get $N_c\sim 320$,
reasonably close to the critical chain length estimated by the
viscosity measurements of Onogi et al\cite{onogi}.  
To the lowest order, the structure factor 
$\hat S(q) = (2/3) \tilde \Gamma(q)$.
It diverges as $|\vec q|^{-2}$ at $y=0$,
analogous to the behavior at a critical point.
The Young's modulus $Y$ of the system is given by\cite{shirish1} 
$Y \propto \hat S(\vec q=0,\omega)$, 
and it will show a dramatic rise as the
chain length $N$ is increased towards $N_c$ for $\mu_0 \ne 0$.
For these reasons we can identify this critical point
with the onset of entanglement.
Note that the approximation to $\hat S(q)$ we have used here will
not work past the transition to entanglement.
There is some indication  of
fluctuating behavior near the entanglement transition in the classic
viscoelasticity measurements of Onogi et al\cite{onogi}, as indicated
by earlier phenomenological analyses\cite{shirish1,shirish2}.

Fluctuations in 
the system tend to decrease as $c_0$ increases away from the critical
point just discussed.
So for finite concentrations we expect mean field theory 
to hold.
In this limit, we get a time-dependent Landau-Ginzburg equation
(time $\equiv~n$), 
by extremizing $\beta {\cal F}$ with respect to $\psi^*$: 

\begin{equation}
\left[ {\partial \over \partial n} - \left({b^2\over 6}\right)
  \nabla^2 - \mu_0 + v \vert \psi(\vec x,n) \vert^2 \right] \psi(\vec
x, n) = 0 
\label{tdgl}
\end{equation}

For the moment, let us suppose that there is no $n$ dependence in
$\psi$.  Then the model reduces to the SCF equation.
Noting the similarity of this equation to the one used to describe
bosonic fluids\cite{fetter}, we seek tube-like solutions in
cylindrical geometry, 
viz., $\psi(\rho,\phi) = \exp(i s \phi) f(\rho)$, $s$ is an integer:

\begin{equation}
{1\over \rho} {d \over d\rho} \left(\rho {df(\rho) \over 
      d\rho}\right) + \left(1- {s^2\over \rho^2}\right) f(\rho) -
  f^3(\rho)=0 
\label{tube}
\end{equation}

where $\rho$ has been made dimensionless by the length scale
 
\begin{equation}
a = b/\sqrt{6 c_0 v}
\label{tuber}
\end{equation}

We see that for $\rho \to 0$,
assuming $|f(\rho)|\to 0$, we get a solution proportional to 
$J_s(\rho) \sim \rho^s$\cite{fetter}.
By choosing the amplitude of this solution correctly, one can match
it for large $\rho$ to a constant solution $f(\rho)=\pm
\sqrt{1-s^2/\rho^2}$ (see Fig.(\ref{fig1})).   
This solution may also be continued mathematically to $s\to 0$. 
But then the $s=0$ solution has an infinite slope at the origin.
The radial extent of these solutions is $\sim s~a$, for $s \ne 0$.
From excluded volume considerations, the maximum density possible in
the system is $1/v$, and for densities approaching $1/v$, $a \sim
b/\sqrt6$.
Physically, these solutions are reminiscent of the tube model of
Edwards\cite{doi-ed}.
They indicate that the tubes are not empty, but have a concentration
of polymers given by $[f(\rho)]^2$.
It may be possible to use numerical simulations to check if the
qualitative features of the occupation profile $[f(\rho)]^2$ can be
reproduced\cite{grest}. 

As discussed by Fetter, the higher the value of $s$, the higher the
energy of the tube-like configuration,
and the system will prefer to have many tubes with a low $s$ than a
few with a large $s$.
We have thus derived from our theory a tube model with a tube radius
$a$, which is 
exceedingly large for small densities, and asymptotes smoothly in an
inverse square root fashion to $\sim b/\sqrt 6$ as the density
approaches $1/v$. 
Since the monomer length $b$ is of the order of a few Angstroms, the
magnitude of the tube radius for $c_0 v \sim 10^{-2}$ is a few
monomer lengths. This is in qualitative agreement with estimates which 
can be found in the literature\cite{doi-ed}.

In the theory of superfluids\cite{fetter}, these cylindrical
structures are viewed 
as idealizations of vortices, caused by the rotation of the fluid.
In our case we
can derive an equation of continuity from Eqn.(\ref{tdgl}):

\begin{eqnarray}
{\partial |\psi|^2 \over \partial n} &&= - \vec \nabla \cdot \vec j + S
\nonumber\\
\vec j &&= \left({b^2\over 6}\right) \left(\psi^* \vec \nabla \psi +
  c.c. \right) \nonumber\\
S &&=  \left({b^2\over 3}\right) \vert \vec \nabla \psi \vert^2 + 2
\mu_0 |\psi|^2 - v |\psi|^4
\label{conti}
\end{eqnarray}

It follows by examining the current $\vec j$, that 
the tube-like 
solutions are not caused by rotation, but rather, there is a {\em radial 
  velocity} due to the $\rho$ dependence of the solution.
There is yet another solution to
Eqn.(\ref{tdgl}), and is obtained by assuming that it is
dependent solely on $n$, which yields
$\psi(n) = [1+\exp(-2 n \alpha)]^{-1/2}$.
More general solutions can undoubtedly be found numerically by
studying Eqn.(\ref{tdgl}) with various boundary conditions. 

We speculate that the phase of the field $\psi$ may
be useful in quantifying the notion of entanglement in polymers.
Ultimately the theory needs to be generalized to handle polymer dynamics.

This research is supported by the Department of Energy contract
W-7405-ENG-36, under the aegis of the Los Alamos National Laboratory
LDRD polymer aging DR program.


%% file: zref.tex

%% file: zfigs.tex
\begin{figure}
\caption{Numerical solutions of Eqn.(\ref{tube}) for s=1 and s=2.
Both curves asymptote to unity, but on different scales.}
\label{fig1}
\end{figure}
\newpage


%% file: zfigps.tex
\begin{figure}

\centerline{Chitanvis,\ \ Fig.\ \ref{fig1}}
\vspace{3cm}
\epsfxsize=12cm
\centerline{\epsffile{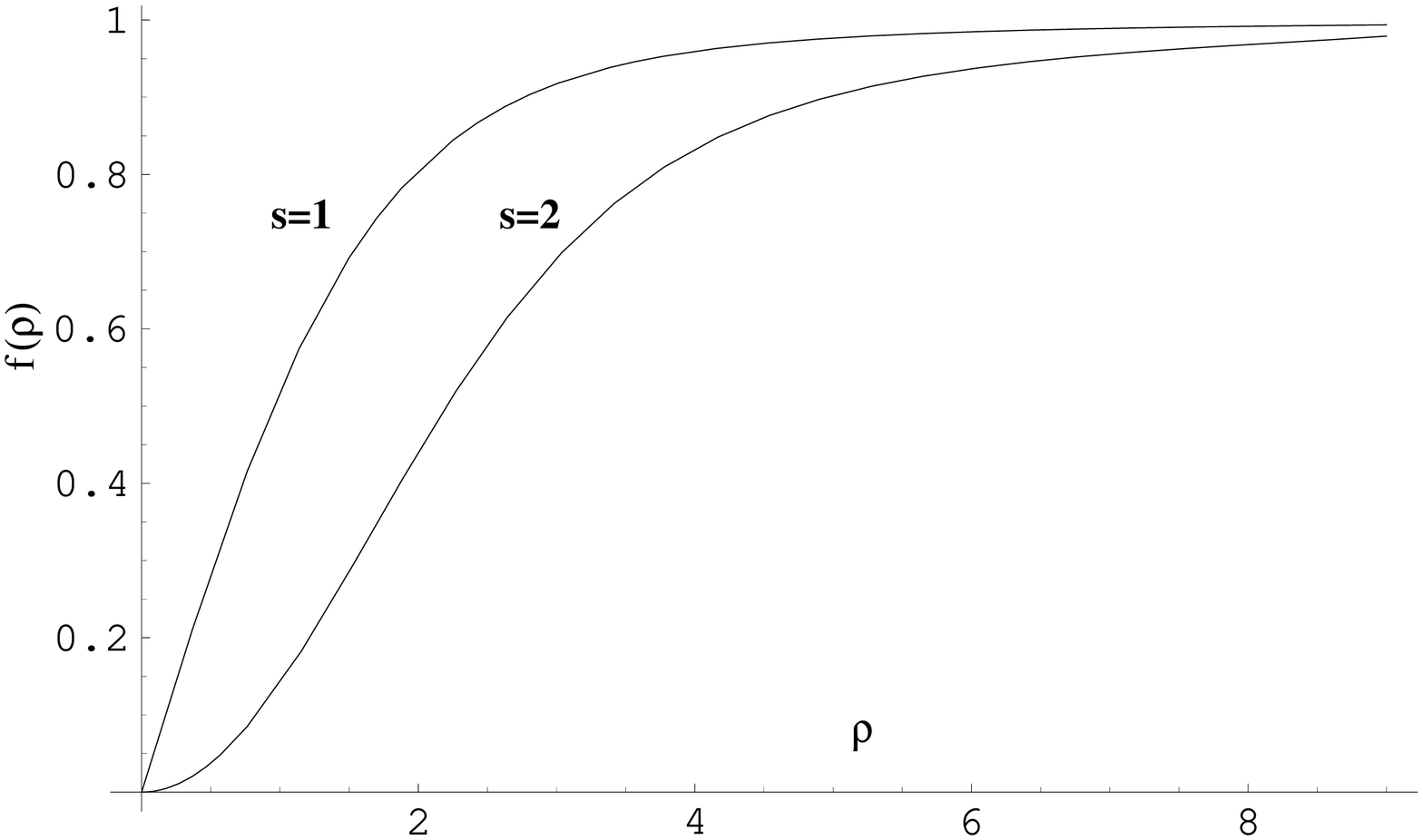}}
\eject
\end{figure}
